\documentclass[11pt,a4paper,nofootinbib]{article} 
 \pdfoutput=1
\usepackage[utf8]{inputenc}
\usepackage{jheppub}
\usepackage{floatflt,rotate, cancel}
\usepackage{mathrsfs}
\usepackage{amssymb, amsmath}
\usepackage{color}
\usepackage{graphicx}
\usepackage{subcaption, caption}
\usepackage{multirow}
\usepackage{float}
\usepackage{pstricks}
 \usepackage{cleveref}
 \usepackage{soul}
\usepackage{jheppub}

\title{Discovery prospects of a light charged Higgs near the fermiophobic region of Type-I 2HDM}

\author{Disha Bhatia$^{\rm a}$, Nishita Desai$^{\rm b}$ and Siddharth Dwivedi$^{\rm c}$}

\affiliation{$^{a}$Instituto de Fisica, Universidade de São Paulo, C.P. 66.318, 05315-970 Sao Paulo, Brazil\\
  $^{b}$Tata Institute of Fundamental Research, Homi Bhabha Road, Colaba,
  Mumbai 400005, India\\$^{c}$Krea University, 5655, Central Expressway, Sri City, Andhra Pradesh, 517646, India}

\emailAdd{disha@if.usp.br}
\emailAdd{nishita.desai@tifr.res.in}
\emailAdd{dwivedi.siddharth@gmail.com}
\abstract{
Determining if the SM-like Higgs is part of an extended Higgs sector is the most important question to be asked after discovery. A light charged Higgs boson with mass smaller than the sum of top and bottom quarks is naturally allowed in Type-I two Higgs doublet model and can be produced in  association with neutral scalars for large parts of parameter space at the LHC. Such low mass charged scalars typically have dominant decays to the fermionic modes viz.\,$\tau\nu$ and $c s$. However in the presence of light neutral scalar ($\varphi$), the charged Higgs boson has a substantial branching fraction into the bosonic decay modes $H^{\pm} \to W^{(*)} \varphi$. Identifying the heavier neutral Higgs ($H$) with the observed 125 GeV Higgs and working in the limit $M_{H^\pm} \approx M_A$, we examine charged Higgs production and decay in the bosonic mode $p p \to H^\pm h \to W^{(*)}h h$.
The presence of two light Higgses ($h$) is then the key to identifying charged Higgs production.  The light Higgs branching ratio is largely dominated by the $b\bar{b}$ mode except when close to the fermiophobic limit.  Here, the rates into $b \bar b$ and $\gamma \gamma$ can be comparable and we can use the $\gamma\gamma b\bar{b}$ signature.  This signature is complementary to the $h h \to 4\gamma$ which has been previously discussed in literature. Using the lepton from the $W$ boson, we demonstrate with a cut-and-count analysis that both the new light neutral Higgs as well as charged Higgs can be probed with reasonable significance at 13.6 TeV LHC with 300-3000 fb$^{-1}$ of integrated luminosity.}


\DeclareUnicodeCharacter{2212}{-}

\begin{document}

\maketitle

\section{Introduction}
Standard Model (SM) of particle physics, even after the discovery of 
the Higgs boson remains far from complete, as there are several compelling experimental evidences and theoretical motivations to believe that physics beyond the Standard Model must exist.
The glaring issues related to the nature of dark matter, neutrino masses and mixings, fermionic mass hierarchy, Higgs mass stability, the number of generations of fermions and the nature of electroweak symmetry breaking are some of the long standing problems which are constantly confronting the SM. 

These problems together hint towards the necessity of having an extended high-energy description of which SM can be a low-energy effective theory. Studying simple extensions of the SM is therefore a reasonable starting point to test against new experimental data.  
One of the most studied extensions include adding scalar fields transforming as multiplets (i.e. $n$-plet) of $SU(2)_L$. Using any $n \geq 2$ predicts new charged particles in addition to multiple new neutral scalars.

In this work, we focus on one such category of well motivated scalar extension of the SM with $n=2$ i.e.\,the Two Higgs doublet model (2HDM)~\cite{Gunion:1989we, Branco:2011iw, Gunion:2002zf, Bhattacharyya:2015nca}.  Different patterns of Yukawa couplings of the two Higgses results in widely different phenomenology and are classified into different types~\cite{Branco:2011iw}.  The Type-I model that we address in this paper is the case where only one of the two Higgs doublets has direct Yukawa couplings to SM fermions.  Spontaneous breaking of the of the electroweak symmetry results in five massive particles viz.\,two neutral CP-even scalars $h$ and $H$ (where one of them has to be identified with the observed 125 GeV scalar), one neutral CP-odd scalar ($A$) and two singly charged scalars ($H^\pm$).  The mixing between neutral scalars is parametrised by the angle $\alpha$ and the ratio of vacuum expectation values of the two higgses by $\tan \beta$.  These two parameters, along with two further mass parameters completely determine the Higgs sector of this model.

These extended sectors with charged Higgs, owing to its rich variety of production and decay topologies have been extensively searched at colliders~\cite{Flechl:2019dtr}. The allowed mass vs. coupling parameter regions of $H^\pm$ obtained from the experiments depend exclusively on the type of 2HDM taken into consideration. For example, mass of $H^\pm$ in Type-II 2HDM is stringently constrained using the experimental data on $B$ decays and on the other hand is very weakly constrained for Type-1 2HDM~\cite{Arbey:2017gmh}.

In this paper, we study the LHC detection prospects of a light charged Higgs boson, specifically, we mean $M_{H^\pm}$ lighter than sum of top and bottom masses i.e. $M_{H^{\pm}} < m_t + m_b$. Generically, a light charged Higgs can be produced at colliders in the following channels: $tb{H^\pm}$, $H^\pm \varphi$, $H^\pm W^\mp$ and, $H^\pm H^\mp$. If kinematically allowed, the produced charged scalar can be detected via its decays to the $W^\pm \varphi$ (where $\varphi = h, H$ or $A$), $\tau{\nu}$, $c s$ and $ t b $.  
The experimental searches of $H^\pm$ have largely focused on the fermionic modes for both its production i.e. $p p \to t H^\pm b$ and decay i.e. $H^\pm \to \tau \nu$ and $c s$~\cite{CMS:2019bfg}. Naively, to perform an experimental search for charged Higgs in fermionic modes is relatively simpler than bosonic modes as bosonic modes require identification of additional intermediate particles along with the charged Higgs.\footnote{For example., when charged Higgs decays to $W \varphi$, then the analysis boils down to first identifying the neutral scalar $\varphi$ in order to accurately estimate the charged Higgs mass.}  This increases the number of unknown parameters in the analysis, making it very sensitive to the mass splittings and couplings under consideration. However, in Type-I 2HDM, couplings of the charged Higgs to fermions scale as $\cot \beta$. Therefore for $\tan\beta \gg 1 $, couplings to fermions are small and fermionic modes are no longer useful probes to search for such a charged Higgs. Therefore in order to fully explore the parameter space, specially the higher $\tan \beta$ regions, the searches in the bosonic modes become paramount~\cite{Akeroyd:1998dt,Arhrib:2016wpw, Arhrib:2017wmo, Arhrib:2020tqk, Bahl:2021str}.  Although there seems to be an apparent disadvantage of having multiple new particles in the final state, the bosonic channels have the advantage of having relatively low, and in some cases almost negligible backgrounds, making them golden modes to look for new physics.

We focus, therefore, on the detection prospects of the light charged Higgs for the case where both its production and decay proceed via bosonic channels, in particular through $H^\pm W^\mp \varphi$ couplings. We choose the dominant associated production channel $H^\pm \varphi$ instead of the SM-like $H^\pm W^\mp$~\cite{Bahl:2021str}.  Therefore, we study the production of charged Higgs in association with an additional neutral scalar $\varphi$. One can further classify various final states depending upon the mass hierarchy of the neutral scalars and the process under consideration\footnote{see for example table 4 of ref.~\cite{Bahl:2021str}}.  For this study, it is the heavier Higgs ($H$) of 2HDM that corresponds to the observed 125 GeV resonance~\footnote{For the opposite case, see for example~\cite{Cheung:2022ndq}.}. In this case, the dominant production cross-section happens in the mode with the lighter, non-SM Higgs, i.e.\,$p p \to H^\pm h$. This is because constraints from electroweak precision observables demand $M_A \approx M_{H^\pm}$ making $p p \to H^\pm A$ cross-section smaller than $p p \to H^\pm h$ due to phase space suppression. Moreover, the production cross-section $p p \to H^\pm H$ is also negligible due to constraints from the measured signal strength of the SM-like Higgs~\cite{CMS:2018uag,ATLAS:2020qdt} which restrict the interactions of the observed Higgs with charged Higgs and imply that the $W^\pm H^\mp H(125)$ coupling is vanishingly small.
\begin{figure}[t!]
    \centering
    \includegraphics[width=0.4\textwidth]{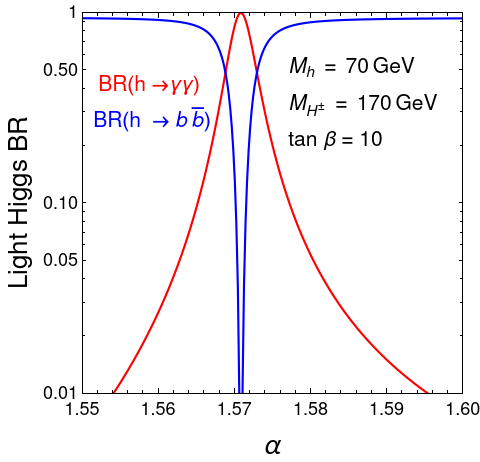}
        \includegraphics[width=0.4\textwidth]{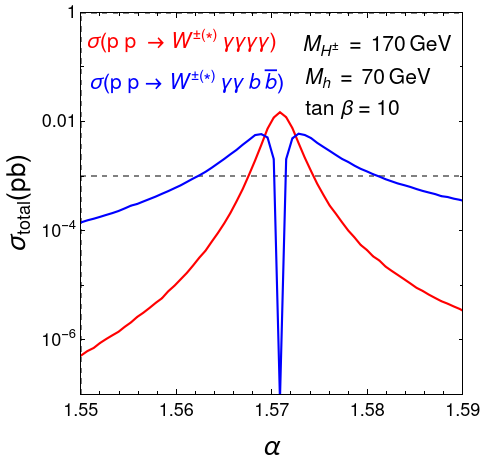}
    \caption{The variation of the light Higgs ($h$) branching ratio (left) and the production cross-section of charged Higgs in the channel $p p \to W^{(*)}h h $ (right) near the fermiophobic limit is shown. For illustration, we use $\tan \beta = 10$, $M_h = 70$ GeV and $M_{H^\pm} = 170$ GeV.}  
    \label{fig:lightHiggsBR}
\end{figure}

In order to probe such a charged Higgs at LHC, we consider its decay to $W^\pm h$.  Thus, the channel under consideration is $p p \to H^\pm h \to {W^{(*)}}^\pm h h $. Reconstructing $H^\pm$ in this case requires identifying two light scalars. Such a light Higgs can decay to a pair of quarks or a pair of photons. The phenomenology of charged Higgs in Type-I 2HDM with both the light Higgses decaying to pair of photons near the fermiophobic regions (resulting in $4\gamma$) has been extensively studied in the literature \cite{Arhrib:2017wmo, Wang:2021pxc}.  However, as can be seen from Fig.~\ref{fig:lightHiggsBR}, even a 1\% deviation from fermiophobic limit ($\alpha = \pi/2$)  makes the light Higgs decay fraction into fermions larger than the di-gamma mode.  We therefore choose the $\gamma \gamma b \bar b$ as our discovery channel.  The presence of the decay to di-gamma reduces SM backgrounds significantly compared to the $4b$ final state. 

Our paper is organised as follows. In  section~\ref{section:theory}, we briefly outline the model, kinematic region of interests and, the parameter space constraints from various theoretical considerations and experimental measurements. In section \ref{section:collider_analysis} we describe the details of our collider analysis, where in section~\ref{subsec:reconst} we 
describe the methodology of tagging multi-body final state which is followed by a discussion of results in section \ref{section:results}. We summarize our findings and conclude in section~\ref{section:conclusions}.

\section{Model description and parameter space constraints}
\label{section:theory}
\subsection{A brief review of 2HDM} \label{subsec:review_2HDM}
As two Higgs doublet models are one of the most well studied beyond standard scenarios in the literature, we shall be very brief in describing the model properties in this
section (for detailed review see ref.~\cite{Gunion:1989we, Branco:2011iw}). Among the four possible CP-conserving 2HDMs, we focus on Type-I 2HDM as a candidate model beyond SM searches. Type-I 2HDM in its construction bears lot of similarities with the SM, specially in the Yuwaka sector. It's Lagrangian is given as: 
\begin{eqnarray}
    \mathcal{L}^{\rm TypeI}_{\rm 2HDM}  &=& D_\mu \Phi_1^\dagger D^\mu \Phi_1 + D_\mu \Phi_2^\dagger \Phi_2 - \bigg(\overline{Q_L} \mathcal{Y}_d \Phi_2 d_R 
                                         +\overline{Q_L} \mathcal{Y}_u \Phi_2^c u_R 
                                         + \overline{Q_L} \mathcal{Y}_e \Phi_2 e_R + h.c. \bigg)  \nonumber \\
&-& m_{11}^2 \Phi_1^\dagger \Phi_1 -\frac{\lambda_1}{2} \left(\Phi_1^\dagger \Phi_1\right)^2 - m_{22}^2 \Phi_2^\dagger \Phi_2 -\frac{\lambda_2}{2} \left(\Phi_2^\dagger \Phi_2\right)^2 + m_{12}^2 \left(\Phi_1^\dagger \Phi_2 + h.c.\right) \nonumber \\
&+& \lambda_3 \left(\Phi_1^\dagger \Phi_1\right) \left(\Phi_2^\dagger \Phi_2\right)
+ \lambda_4 \left( \Phi_1^\dagger \Phi_2 \right) \left( \Phi_2^\dagger \Phi_1\right) 
+ \frac{\lambda_5}{2} \left(\left( \Phi_1^\dagger \Phi_2\right)^2 + h.c.\right) \;.
\label{eqn:2HDM}
\end{eqnarray}

Spontaneous breaking of the electroweak symmetry relates the couplings in the scalar potential with the physical mass parameters of the five scalar fields and the mixing angles --- $\alpha$ (related to diagonalization of CP-even Higgses) and $\beta$ (related to diagonalization of the charged as well as the CP-odd Higgses). We list some of the important couplings of the physical states which are useful in our analysis. 
The bosonic part of the Lagrangian includes:
\begin{align}
    \mathcal{L}_{\rm bosons} & \supset 
    \frac{2 M_W^2}{v}  \sin(\beta-\alpha) h W_\mu^+ {W^{\mu}}^{-} 
    + \frac{M_Z^2}{v}  \sin(\beta-\alpha) h Z_\mu Z^\mu \nonumber \\
    & + 
     \frac{2 M_W^2}{v} \cos(\beta-\alpha) H W_\mu^+ {W^\mu}^{-} 
    + \frac{M_Z^2}{v} \cos(\beta-\alpha) H Z_\mu Z^\mu  \nonumber \\
    &-  \frac{g}{2} \cos(\beta-\alpha) \left[ W^+_\mu \left(h \;\partial^\mu H^-  
    - H^{-} \;\partial^\mu h \right) - W^-_\mu \left(h \; \partial^\mu H^+  
    - H^{+} \; \partial^\mu h \right) \right]\nonumber \\ 
    &+  \frac{g}{2} \sin(\beta-\alpha) \left[  W^+_\mu \left(H \; \partial^\mu H^- - H^{-} \; \partial^\mu H \right) - W^+_\mu \left(H \; \partial^\mu H^+  - H^+ \; \partial^\mu H \right)  \right]\nonumber \\
    &- i \frac{g}{2} W^+_\mu \left[A \; \partial^\mu H^-   
    - H^{-} \; \partial^\mu A \right] + h.c. \;
    \label{eqn:bosons}
\intertext{and the fermionic part of the Lagrangian includes}
    \mathcal{L}_{\rm fermions} &\supset 
    - \frac{\cos\alpha}{\sin\beta}\sum_f \frac{m_f}{v} \overline{f}f h 
     -  \frac{\sin\alpha}{\sin\beta}\sum_f \frac{m_f}{v} \overline{f}f H 
     \nonumber \\
    & - \frac{\sqrt{2}}{v}\cot\beta \left( \sum_u V_{ud} \overline{u} \left( m_u P_L + m_d P_R \right)d  
    + \sum_\ell m_\ell \overline{v_\ell}  P_R \ell\right) \; H^+
    + h.c. \;. 
    \label{eqn:fermions}
\end{align}
\subsection{Considerations for experimental discovery}
We can make following inferences from the above Lagrangian: 
\begin{enumerate}
    \item The limit $\beta \to \alpha$ corresponds to the alignment limit
    as the couplings of $H$, which in our case represents the observed 125 GeV Higgs, become 
    SM-like. Since no new sign of new physics is indicated from the Higgs 
    coupling measurements~\cite{CMS:2018uag, ATLAS:2020qdt}, any 
    deviation away from the alignment limit is heavily constrained.

\item Charged Higgs couplings to fermions scale as $\cot \beta$, therefore for $\tan \beta \gg 1$, these channels become insensitive at colliders. Consequently, in order to constrain higher $\tan \beta$ regions in Type-I 2HDM, the experimental searches must probe the charged Higgs boson in the bosonic channels. 

 \item The simplest bosonic decay of charged Higgs is $H^\pm \to W^\pm H$ as no further new beyond-SM particles are produced. However the Higgs signal strength measurements~\cite{CMS:2018uag, ATLAS:2020qdt} demand Higgs coupling to be nearly aligned with the SM thereby restricting the $H^\pm W^\mp H$ coupling. This makes the branching fraction of charged Higgs in the $W^{(*)}H$ channel negiligible. Therefore, in order to probe Type-I charged Higgs at colliders, one has to rely on its production in association with additional scalar particles.

\item The light Higgs in Type-I 2HDM possesses an interesting limit where its decays to fermions ($f\bar{f}$) become negligible, and for a sufficiently light $h$ (where decays to di-bosons $WW^*$ and $ZZ^*$ are also phase space suppressed), the diphoton mode dominates over others despite it being loop suppressed \footnote{See ref. \cite{Gunion:1989we, Barger:1993my,Kniehl:1995tn, Djouadi:2005gj} for one-loop form factors for $h\gamma \gamma$ coupling}.  This limit is known as fermiophobic limit and occurs when $\alpha \to \pi/2 $.  Since QCD background is much reduced in the diphoton channel, it seems quite favourable to probe light Higgs in regions near the fermiophobic limit.  The pseudoscalar ($A$) does not have a fermiophobic limit and its decays into $f \bar{f}$ channels are always allowed whenever kinematically accessible.

\end{enumerate}

Using the properties of the scalars couplings in Type-I 2HDM mentioned in section~\ref{subsec:review_2HDM}, we describe the best channel for probing such a charged Higgs at the LHC.  Searches for $H^\pm$ have been extensively performed in the fermionic modes \cite{CMS:2015lsf, ATLAS:2016avi, CMS:2019bfg, CMS:2020imj, ATLAS:2021upq, Arbey:2017gmh}. However, the fermionic modes get suppressed for regions with $\tan \beta \gg 1$. Therefore an independent analysis with bosonic modes is quintessential for probing the parameter space of Type-1 2HDM. 
There are three main channels for the production of charged Higgs in the bosonic modes at the colliders viz.\,$H^{\pm} \varphi$ (where $\varphi=h, A$), $H^{\pm} W$ and $H^{+} H^{-}$. We choose the charged Higgs production in association with the scalar $\varphi$ owing to its larger cross-section in comparison to the rest~\cite{Bahl:2021str}.
In our set-up, for simplicity, we also consider the masses of pseudoscalar and the charged Higgs scalar to be equal. This is well motivated by the $\rho$ parameter constraint~\cite{PhysRevD.46.381, Gunion:1989we, Workman:2022ynf} of electroweak precision tests. Since $M_h < M_{H^\pm} = M_A$, the production cross-section for $H^\pm h$ naturally becomes larger than $H^\pm A$. Furthermore, when kinematically allowed, the charged Higgs decay fraction to $W^\pm h$ is also substantial. Therefore, we choose $p p \to H^\pm h \to W^{(*)} h h$ as our channel to probe charged Higgs at the colliders. In order to reconstruct the charged Higgs, we require to efficiently tag both the light Higgses. The dominant decays of light Higgs is in the $b\bar{b}$ channel over large parts of parameter space. However probing $hh \to 4 b$ at LHC is challenging because of large QCD backgrounds.
Interestingly, in Type-1 2HDM, there is a limit as discussed in section~\ref{subsec:review_2HDM} where the branching ratio to fermions vanishes. 
In this limit, the analysis of $hh\to 4 \gamma$ has been performed in ref.\cite{Arhrib:2017wmo, Wang:2021pxc}. However the branching ratio to diphoton falls off very sharply away from the fermiophobic regions (see figure~\ref{fig:lightHiggsBR}).  Consequently, branching into the dominant fermionic mode ($b \bar{b}$) also quickly rises.  Therefore, in order to probe regions close to $\alpha \to \pi/2$, we tag the di-Higgs
in the $\gamma\gamma b\bar{b}$ mode. Our analysis therefore serves as a complementary probe to the 
analysis in ref.~\cite{Arhrib:2017wmo, Wang:2021pxc} which can also be seen from the right panel of figure~\ref{fig:lightHiggsBR}.

\subsection{Parameter space scan and constraints}
\label{subsec:param_space}

\begin{figure}[t]
    \centering
    \begin{subfigure}{0.48\textwidth}
    \centering
    \includegraphics[width=\textwidth]{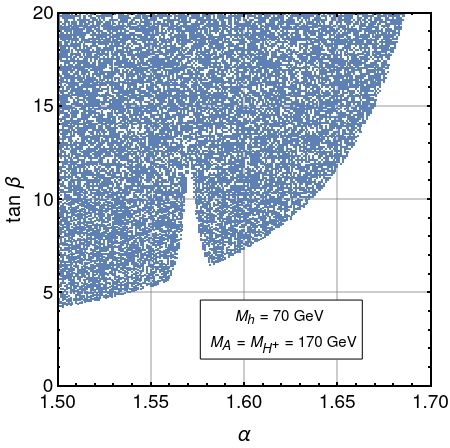}
    \caption{}  \label{subfig:param_mh_70}
    \end{subfigure}
    \hfill
      \begin{subfigure}{0.48\textwidth}
    \includegraphics[width=\textwidth]{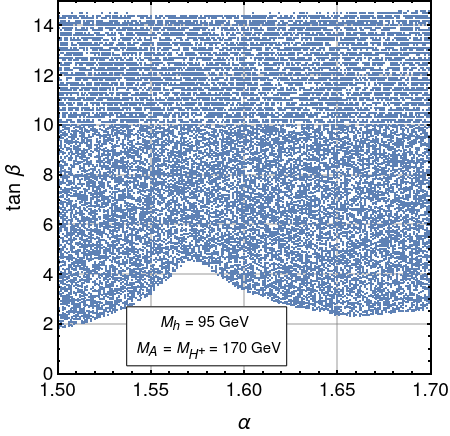}
    \caption{} \label{subfig:param_mh_95}
     \end{subfigure}
     \begin{subfigure} {0.48\textwidth}
    \includegraphics[width=\textwidth]{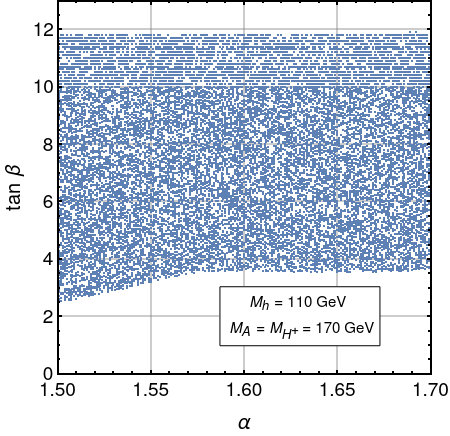}
    \caption{}  
    \label{subfig:param_mh_110}
    \end{subfigure}
    \caption{Allowed region in the   $\alpha$ vs $\tan \beta$ plane for  $M_h = 70$ GeV (\ref{subfig:param_mh_70}), $M_h = 95$ GeV (\ref{subfig:param_mh_95}) and $M_h = 110$ GeV (\ref{subfig:param_mh_110}) respectively. For all panels $M_{H^\pm} = M_A = 170$ GeV. }
    \label{fig:param_space}
\end{figure}

The two Higgs doublet model described in eq.~\ref{eqn:2HDM} 
can be described in terms of the physical masses of the scalars, 
mixing angles and the $Z_2$ symmetry breaking parameter $m_{12}$.
The identification of $M_{H^\pm} = M_A$ and $M_H = 125$ GeV, 
leaves a total of five independent parameters viz.\,$M_h$, $M_{H^\pm}= M_A$, 
$m_{12}$, $\alpha$ and $\tan \beta$~\cite{Branco:2011iw}. In order to determine 
the allowed parameter space taking various theoretical and experimental constraints into account, 
we vary these parameters in the following ranges:
\begin{enumerate}
    \item \textbf{Light Higgs mass}: The light Higgs mass lies in the range [65, 110] GeV. The choice is made to avoid the bounds from the observed Higgs decays $H \to hh$~\cite{CMS:2018uag, ATLAS:2020qdt}. 
    \item \textbf{Charged Higgs mass}: We place an upper limit from the requirement that the charged Higgs must be smaller than the $t\bar{b}$ threshold, while the lower limit on $M_{H^{\pm}}$ is an experimental bound from LEP constraining its mass to be greater than approximately 80 GeV \cite{ALEPH:2013htx}. However, since here we consider the decays of charged Higgs to $W^{(*)}h$, the mass range of interest is around $M_{H^\pm} \in [140, 170]$ GeV in order to avoid large phase space suppression. As a conservative illustration, we consider a maximal value of $M_{H^\pm} = 170$ so that the $Wh$ decay is always on-shell for the entire range of light Higgs masses.
    \item \textbf{The mass parameter $m_{12}$}: We vary this parameter in the range [1,100] GeV. 
    However, even though it does not directly affect the signal production, it can limit allowed values of $\tan \beta$. We fix it appropriately to a suitable value to allow maximum variation in $\tan \beta$. For more details see discussion in appendix.~\ref{append:m_12_values}.
    \item {\boldmath${\alpha}$}: The mixing angle is varied in the full range [0, $\pi$] to determine constraints 
    on parameter space. For collider signatures, we choose regions close to the fermiophobic limit {\it i.e.} $\alpha \in [1.5, 1.7]$. 
    \item \boldmath{$\tan \beta$}: The ratio of Higgs vacuum expectation values is varied between [1, 20].  
\end{enumerate}

To determine the allowed range of parameters, we randomly sample points in this space and apply the following theoretical and experimental constraints on our benchmark sets.   
\begin{itemize}
 \item The quartic couplings of the scalar potential of 2HDM can be expressed in terms of physical masses and mixing angles~\cite{Branco:2011iw}. We ensure that the chosen points satisfy vacuum stability, perturbativity and unitarity constraints~\cite{PhysRevD.18.2574,Kanemura:1993hm, Akeroyd:2000wc, Branco:2011iw}.
 \item The constraints from $\rho$ parameter which relates the masses 
 of the additional scalars $M_h$, $M_{H^\pm}$ and $M_A$~\cite{ Gunion:1989we, Workman:2022ynf} are also imposed. 
 These can be satisfied when $M_{H^\pm} \approx M_A$ and we work in the limit  $M_{H^\pm} = M_A$ for this work. 
  \item Another set of constraints arise from the identification of the heavier neutral Higgs of Type-I 2HDM with the observed 125 GeV Higgs. Since Higgs couplings are modified by functions of 2HDM parameters (see eq.~\ref{eqn:bosons},\ref{eqn:fermions}), we can get strong constraints from the signal strength measurements \cite{CMS:2018uag,ATLAS:2020qdt}.
  \item The direct searches of additional scalars also restrict the allowed parameter space further. 
   We impose the constraints from LEP, and Tevatron~\cite{LEPHiggsWorkingGroup:2001opo, ALEPH:2006tnd, PhysRevD.87.052008, refId0}, as well as the low mass diphoton resonance searches from ATLAS and CMS~\cite{ATLAS-CONF-2018-025, CMS:2018cyk}. Other direct searches at the LHC mostly concern regions with heavy charged Higgs and pseudoscalar boson and do not directly apply to our parameter space~\cite{ATLAS:2017ayi, ATLAS:2018gfm, CMS:2019qcx, CMS:2019ogx}. Even for the regions where there is some overlap with the our chosen 
   range, we find that the resulting constraints do not rule out any additional parameter space. 
\end{itemize}

After applying all the above constraints, the number of points we are left with is shown in \ref{tab:benchmarks} along with chosen benchmark parameters.
In figure~\ref{fig:param_space}, we present our results in the plane of $\left( \alpha, \tan \beta \right)$ for these three benchmark cases listed in Table~\ref{tab:benchmarks}.  We find that applying theoretical and experimental constraints reduces allowed parameter space by nearly 50\%, but still significant regions remain unprobed.

\begin{table}[t!]
\begin{center}
\begin{tabular}{|c|c|c|c|c|}
\hline
 $M_h$ & $M_{H^\pm}$  & $m_{12}$ & $N_\mathrm{start}$  &  $N_\mathrm{final}$  \\ \hline
70 GeV & 170 GeV  & 15 GeV & 50000  & $\approx 28000$  \\ \hline
 95 GeV & 170 GeV  & 25 GeV & 50000  &$\approx 30000$    \\ \hline
110 GeV & 170 GeV  & 32 GeV & 50000  & $\approx 22000$   \\ \hline
\end{tabular}
\end{center}
\caption{Parameter scan benchmarks for determining allowed parameter space.  In all cases, we scan $\tan \beta$ between 1--20 and $\alpha$ between $0-\pi$. The number of points in the scan and the number that survive all known constraints are also shown. The choice of $m_{12}$ is explained in the appendix~\ref{append:m_12_values}. 
\label{tab:benchmarks}}
\end{table}
\begin{itemize}
    \item Theoretical constraints in general are weaker than the experimental 
    constraints. However the appearance of the upper bound on $\tan \beta$ 
    for $M_h = $ 95 and 110 GeV arises 
    due to the positivity requirement $\lambda_1 > 0$ on the quartic coupling (see eq.~\ref{eqn:lambda1})
    of the scalar potential.
    \item The wedge shaped disallowed regions in the figure~\ref{fig:param_space} arise due to low mass diphoton 
    constraints from ATLAS and CMS \cite{ATLAS-CONF-2018-025, CMS:2018cyk}. The exclusion is more prominent for 
    $M_h = 70$ and 95 GeV than for $M_h= $ 110 GeV because of more stringent constraints on low mass Higgs. 
    \item The dominant LEP constraint arises from the $Z h$ production \cite{LEPHiggsWorkingGroup:2001opo} 
    which significantly rules out the parameter space  
    specially for the lower masses of light Higgs (see figure~\ref{fig:param_space} for $M_h = $ 70 GeV). 
    For $M_h = 95$ and 110 GeV, the bounds are weaker because of phase space suppression at LEP due to limited available center of mass energy. 
    \item Higgs signal strength measurements do not provide further constraints on our parameter space beyond the ones coming from direct searches of the light Higgs~\cite{LEPHiggsWorkingGroup:2001opo,ATLAS-CONF-2018-025, CMS:2018cyk}.
\end{itemize}

\section{Collider signatures}
\label{section:collider_analysis}

In this section, we describe the numerical framework to probe the allowed parameter space in figure~\ref{fig:param_space} 
at 13.6 TeV LHC in the channel $p p \to W^{(*)} h h$. To minimize QCD backgrounds, we tag the $W$ boson through its leptonic decays to $\ell {\nu_\ell}$ ( $\ell = e$ or $\mu$). The light Higgs in large parts of the parameter space decays dominantly to $b\overline{b}$ excepting the regions around the fermiophobic limit where the decays to diphotons become substantial. For the above channel, 
both light Higgses decaying via $h \to \gamma \gamma$ giving rise to the $W + 4 \gamma $ final state has been considered in ref. \cite{Arhrib:2017wmo,Wang:2021pxc}. 
Here, we consider one of the light Higgs decaying to $b \bar b$ and other to $\gamma \gamma$. This channel serves as a complementary probe to the $W+4\gamma$ final state (see figure~\ref{fig:lightHiggsBR}). 

The dominant background contribution comes from the following processes $p p \to t\bar{t} j \gamma$ and $p p \to t\bar{t} \gamma \gamma$.
The irreducible background  $p p \to W bb \gamma \gamma$ arises due to a combination of mixed QCD-QED and pure QED diagrams, along with resonant Z and SM Higgs contributions. However, the irreducible component does not contribute significantly to the total background and for the analysis that follows we neglect this component. 

 In addition to the signal channel considered here, the same final state can also arise from the process $p p \to H^{\pm} A \to {W^{\pm}}^{(*)} h A$ with $h \to \gamma \gamma$, $A \to b\bar{b}$ and vice-versa. However since $M_h < M_A$, the $H^{\pm} A$  process is phase space suppressed and can be safely neglected for the chosen benchmarks.

To simulate the events for our analysis,  we first implement Type-I 2HDM Lagrangian in {\tt FeynRules}~\cite{Alloul:2013bka}
to generate a Universal FeynRules Output (UFO). The UFO file is interfaced with  
$\texttt{MadGraph5\_aMC@NLO}$~\cite{Alwall:2014hca}  to generate the parton level signal and backgrounds at the leading order\,(LO).
In { \texttt MadGraph}, we apply following basic minimal cuts at the event generation level on the transverse momenta ($p_T$), pseudo-rapidity ($\eta$) and, the angular separation between two final state particles ($\Delta R_{ab}$) keeping in mind the detector's acceptance and triggers:
\begin{eqnarray}
&&{p^{\ell,\; \gamma}_{T}} \geq 10~ {\rm GeV} \;, |\eta_{\ell, \; \gamma} | \leq 2.5 \;, 
\Delta R_{\ell \gamma \;, \gamma \gamma} \geq 0.4  \;, \Delta R_{\ell \ell} \geq 0.2 \;, \nonumber \\
&&{p^{j,\; b}_{T}} \geq 20 ~{\rm GeV} \;, |\eta_{b, ~j} |\leq 5 \;,
\Delta R_{(bb, ~jj, ~j\gamma, ~b\gamma,~\ell j, ~\ell b) } \geq 0.4 \;.
\label{eqn:cutsmad}
\end{eqnarray}
The parton level events after application of above cuts in eq.~\ref{eqn:cutsmad} are then passed
through {\tt PYTHIA 8}~\cite{Sjostrand:2014zea} for showering and hadronization. 
The detector simulation of events is performed using the 
the fast-sim package {\tt Delphes3}~\cite{deFavereau:2013fsa} in   
which we have used the default CMS card for our analysis. 
Final state hadrons are clustered using anti-kT algorithm~\cite{Cacciari:2008gp}
into jets employing the {\tt FastJet} package~\cite{Cacciari:2011ma}.

\subsection{Signal region selection and kinematic variables}
\label{subsec:reconst}

\begin{figure}[ht]
    \centering
    \begin{subfigure}{0.48\textwidth}
    \centering
    \includegraphics[width=\textwidth]{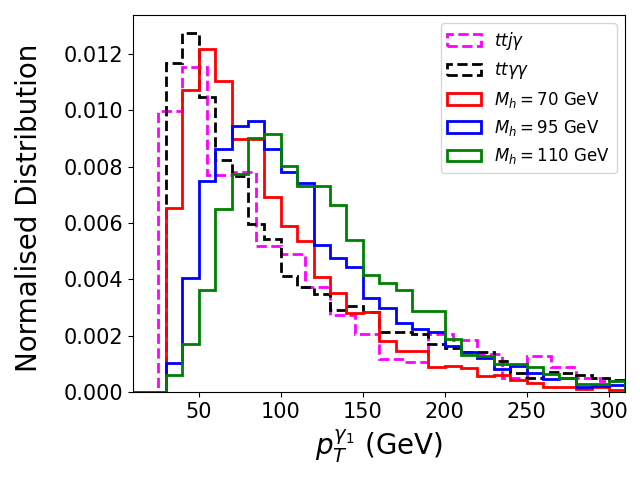}
    \caption{}  \label{subfig:pt_gamma1_select}
    \end{subfigure}
   \hfill
      \begin{subfigure}{0.48\textwidth}
    \includegraphics[width=\textwidth]{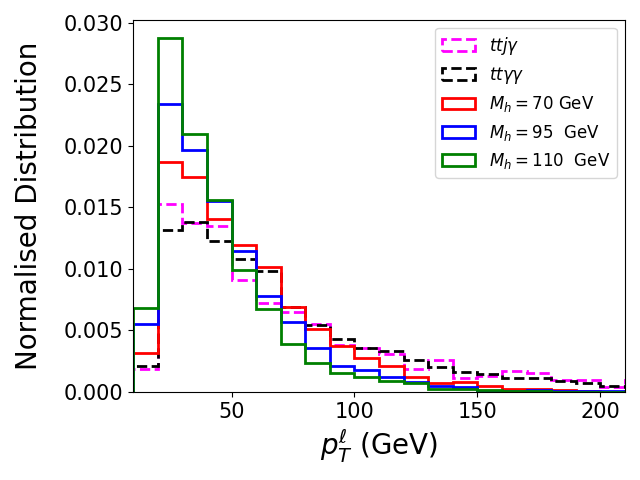}
    \caption{} \label{subfig:pt_lep_select}
     \end{subfigure}
      \begin{subfigure}{0.48\textwidth}
    \includegraphics[width=\textwidth]{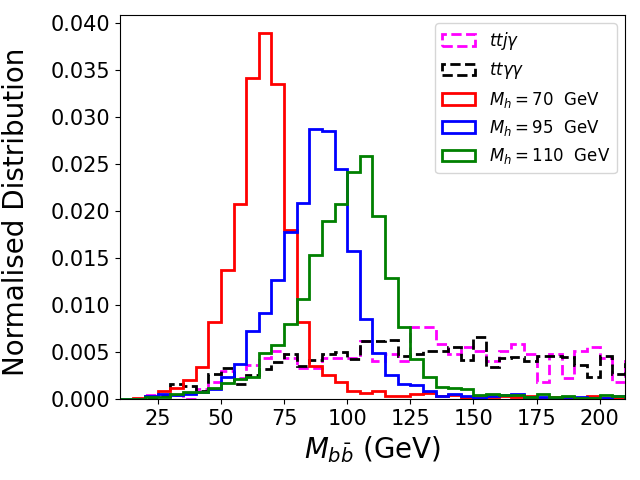}
    \caption{} \label{subfig:M_bb_select}
     \end{subfigure}
     \hfill
     \begin{subfigure} {0.48\textwidth}
    \includegraphics[width=\textwidth]{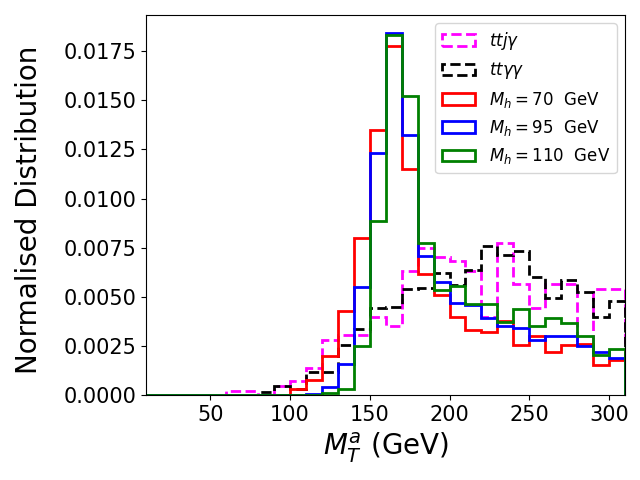}
    \caption{}  
    \label{subfig:MT_total_select}
    \end{subfigure}
      \begin{subfigure} {0.48\textwidth}
    \includegraphics[width=\textwidth]{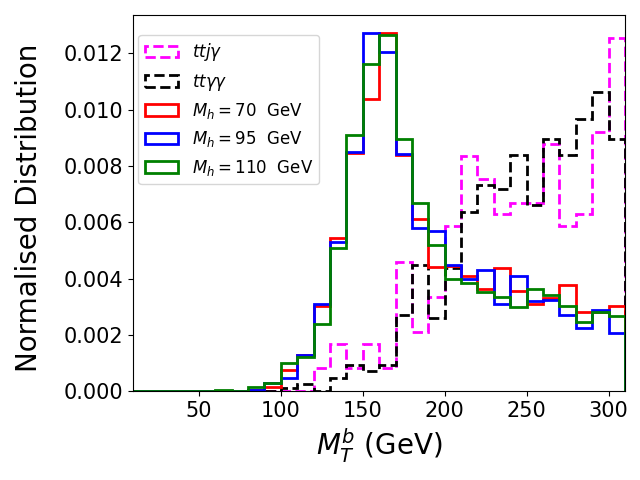}
    \caption{}  
    \label{subfig:MT_total_2_select}
    \end{subfigure}
    \caption{Normalised distributions of the kinematic variables made from the $\ell \gamma \gamma b \bar b + \cancel E_T$ final state after preselection.  We show (a) $p^{\gamma_1}_{T}$, (b) $p^{\ell}_{T}$, (c) $M_{b\bar b}$, (d) $M^{a}_{T}$ and (e) $M^b_T$ as described in section~\ref{subsec:reconst}. }
    \label{fig:variables}
\end{figure}

 To improve the signal to background ratio, we employ the selection criteria for the final state in three steps. We first impose the pre-selection criteria, which helps in the significant reduction of the background. The second stage of selection helps to tag both the light Higgses in the $\gamma \gamma$ as well as $b \bar b$ mode in the mass range 65--110 GeV. In the final stage we aim the selection to reconstruct the charged Higgs. We describe these steps below:
\begin{enumerate}
 \item {\bf Pre-selection:} We demand our final state to have exactly one isolated lepton ($\mu$ or $e$), at least two isolated photons and, two $b-$jets along with missing energy $\cancel E_T$ with the following thresholds \cite{CMS:2018cyk}:
 \begin{align}
  {p^{\ell}_{T}} \geq 18 ~{\rm GeV}, ~~{p^{ \gamma_1}_{T}} \geq 30 ~{\rm GeV}, ~~ {p^{ \gamma_2}_{T}} \geq 18 ~{\rm GeV}~ {\rm and }~ \cancel E_T \geq 20 ~{\rm GeV} \;.
 \end{align}
 \item {\bf  Di-Higgs selection:} 
 In our analysis, we tag one of the light Higgses in the diphoton mode and the another in $b\overline{b}$
 mode. We first describe the light Higgs tagging to diphoton. 
 This part of the analysis is motivated by the low mass resonance ATLAS/CMS searches~\cite{ATLAS-CONF-2018-025,CMS:2018cyk}. Since the transverse momentum spectrum of the diphoton system depends on the mass of the light Higgs, we employ mass specific cuts for $p_T$ of the leading photon. This can be seen from the normalised kinematic distributions shown in figure ~\ref{fig:variables}. Here, we have plotted the transverse momentum of the leading photon and the lepton for background channels, and signal processes corresponding to three benchmark values $M_h = 70, \; 95$ and 110 GeV. 
 Note that the case with $M_h = 95$ GeV is also special owing to a small excess observed by CMS in the diphoton mode for $M_h \sim 95.3$ GeV with a local significance of 2.8$\sigma$ \cite{CMS:2018cyk}. Since the peak of the lepton transverse momentum distribution  does not shift with the mass of the light Higgs, we do not impose any further cut on the lepton $p_T$ after the pre-selection. The cuts on the leading photon $p_T$ for our three benchmark cases are taken as:
 \begin{eqnarray}
  {p^{ \gamma_1}_{T}} &\geq& 40 ~{\rm GeV} ~({\rm for}~ M_h = 70,~95 ~\rm GeV)\nonumber \\
   {p^{ \gamma_1}_{T}} &\geq& 60 ~{\rm GeV} ~({\rm for}~ M_h = 110 ~\rm GeV) \;.
  \end{eqnarray}

  In addition to the $p^{ \gamma_1}_{T}$ thresholds, we also require the invariant mass of the diphoton pair $M_{\gamma \gamma}$ to satisfy
  \begin{align}
        65 < M_{\gamma \gamma} < 115~ \text{GeV} \;.
  \end{align}

 Now we consider the other light Higgs decay to $b\overline{b}$. Apart from a basic
 $p_T$ cut at the level of pre-selection we require the invariant mass of the 
$b \bar b$ system ($M_{bb}$) to lie within
  \begin{align}
      60 < M_{b \bar b} < 115 ~\text{GeV}\;.
  \end{align}

  \item {\bf Reconstruction of the charged Higgs :} 
  The final step of our analysis is to reconstruct the charged Higgs boson. Due to the presence of missing energy in the leptonic decays of the $W$ boson, we cannot fully reconstruct the mass of the charged Higgs boson. Thus, we use a cluster transverse mass variable \cite{Han:2005mu}, whose Jacobian peak corresponds to the true value of $M_{H^\pm}$. Among the two light Higgses in the final state, the one coming from of the decay of the charged Higgs i.e. $H^\pm \to W^{(*)} h$ would correctly reconstruct its mass as opposed to one which is produced in association i.e. $p p \to H^\pm h$. 
  Furthermore, the light Higgs arising from the decay of charged Higgs could decay either to $\gamma \gamma$ or to $b \bar{b}$. 
  To resolve the ambiguity, we first define the transverse mass using both combinations of final decay states of charged Higgs i.e. $\ell\gamma\gamma \cancel E_T$ and $\ell b \bar b \cancel E_T$ system i.e.
\begin{eqnarray}
M^a_T = M_T(\ell\gamma\gamma \cancel E_T) &=& \left(\sqrt{\Vec{p}_{T, \ell \gamma \gamma} + M^2_{\ell \gamma \gamma} + \cancel{E}_T} \right)^2  - \left(\Vec{p}_{T, \ell \gamma \gamma } + \cancel{E}_T \right)^2 \;, \nonumber \\
M^b_T = M_T(\ell b\bar{b} \cancel E_T) &=& \left(\sqrt{\Vec{p}_{T, \ell b \bar b} + M^2_{\ell b \bar b} + \cancel{E}_T} \right)^2  - \left(\Vec{p}_{T, \ell b \bar b} + \cancel{E}_T \right)^2 \;.
 \label{Eq:MT_variable_def}
\end{eqnarray} 
Here $\Vec{p}_{T, \ell \gamma \gamma (\ell b \bar b)}$ denotes the transverse momentum of the $\ell \gamma \gamma (\ell b \bar b)$ system, $M_{\ell \gamma\gamma ( \ell b\bar{b})}$
denotes the invariant mass of the visible component in the $H^{\pm}$ decay and $M^{a(b)}_T = M_T (\ell \gamma\gamma (\ell b\bar{b}) \cancel E_T)$ corresponds to the cluster transverse mass of the $\ell \gamma \gamma (\ell b \bar b) \cancel E_T$ system respectively. The distributions corresponding to $M^a_T$ and $M^b_T$ after preselection cuts are shown in figure \ref{subfig:MT_total_select} and \ref{subfig:MT_total_2_select} respectively.

\noindent Since in a given signal event either the diphoton or the $b \bar b$ pair can come from $H^\pm \to W^{(*)} h$ decay, we demand 
\begin{eqnarray}
&&  |M^a_T - M_{H^\pm} |< \Delta  ~ \text{with} ~(M^b_T - M_{H^\pm} > \Delta ~ \text{or}~ M^b_T - M_{H^\pm}< - \Delta )\nonumber \\
&& \text{or} \nonumber \\
&&  |M^b_T - M_{H^\pm}| < \Delta  ~ \text{with} ~ (M^a_T - M_{H^\pm} > \Delta ~ \text{or}~ M^a_T - M_{H^\pm} < - \Delta ) \;.
  \label{eqn:chargedhiggsselection}
\end{eqnarray}
 The above criterion just ensures that in a given event only one of $\ell\gamma\gamma \cancel E_T$ or $\ell b\bar{b} \cancel E_T$ systems correctly reconstructs $M_{H^\pm}$ and we conservatively select events in a 20 GeV bin (with $\Delta = 10$ GeV) around the true value of the charged Higgs mass. 
\end{enumerate}

We impose these cuts on our benchmark points and backgrounds to estimate efficiencies.  The signal cross-section varies as a function of the independent parameters in Type-1 framework --- not just the masses, but also the mixing parameters $\alpha$ and $\beta$.  
 On the other hand, the change due to variation in mixing angles is only in the overall scale factor and the shape the normalised distributions remains approximately the same.  Therefore, we can determine the signal efficiency for a given benchmark by appropriately fixing $\alpha$ and $\beta$.  
For estimating efficiencies in Table~\ref{table:cut_flow_efficiency}, we fix both the angles equal to $\pi/4$ which make the coupling modifiers either one or zero.

\subsection{Results} \label{section:results}

\begin{table}[]
\begin{tabular}{|c|c|c|c|c|c|}
\hline
Process & Pre. & $M_{\gamma \gamma}$ & $M_{\gamma\gamma} + M_{b\bar{b}}$ & $M_T^a(M_T^b)$ & $N_\mathrm{events}$ \\ 
& & & & & ($\mathcal{L} = 10^3/$fb) \\
\hline
\begin{tabular}[c]{@{}c@{}}$p p \to ttj \gamma$ \\ ($N_\mathrm{MC} = 941071$)\\ $\sigma = 310.7 $ fb\end{tabular}   &  \begin{tabular}[c]{@{}c@{}} $7.54$ \\ $\times 10^{-4}$ \end{tabular} & \begin{tabular}[c]{@{}c@{}}$2.14 \times 10^{-4}$\\ ($1.41 \times 10^{-4}$)\end{tabular}         & \begin{tabular}[c]{@{}c@{}}$4.57 \times 10^{-5}$\\ ($ 3.10 \times 10^{-5}$)\end{tabular}                           & \begin{tabular}[c]{@{}c@{}}$6.38 \times 10^{-6}$\\ ($3.20 \times 10^{-6}$)\end{tabular}          & \begin{tabular}[c]{@{}c@{}}1.98\\          (0.99)\end{tabular}                          \\ \hline 
\begin{tabular}[c]{@{}c@{}}$p p \to tt\gamma\gamma$\\  ($N_\mathrm{MC} = 50$K)\\ $\sigma = 1.403 $ fb\end{tabular} & 0.0474                & \begin{tabular}[c]{@{}c@{}}0.0135 \\          (0.0092)\end{tabular}                             & \begin{tabular}[c]{@{}c@{}}0.0026 \\ (0.0019)\end{tabular}                                                         & \begin{tabular}[c]{@{}c@{}}0.0002 \\ (0.00012)\end{tabular}                                      & \begin{tabular}[c]{@{}c@{}}(0.28)\\          (0.17)\end{tabular}                        \\ \hline \hline
\begin{tabular}[c]{@{}c@{}}$M_h = 70$ GeV\\ ($N_\mathrm{MC} =50$K)\\ $\sigma_0 = 1.0 $ fb\end{tabular}            & 0.0730                & 0.0676                                                                                          & 0.0482                                                                                                             & 0.0207                                                                                           & 20.7                                                                                    \\ \hline
\begin{tabular}[c]{@{}c@{}}$M_h = 95$ GeV\\ ($N_\mathrm{MC} = 50$K)\\ $\sigma_0 = 1.0 $ fb\end{tabular}            & 0.0756                & 0.0741                                                                                          & 0.0657                                                                                                             & 0.0258                                                                                           & 25.8                                                                                    \\ \hline
\begin{tabular}[c]{@{}c@{}}$M_h = 110$ GeV\\ ($N_\mathrm{MC} = 50$K)\\ $\sigma_0 = 0.1 $ fb\end{tabular}           & 0.0721                & 0.0667                                                                                          & 0.0517                                                                                                             & 0.0206                                                                                           & 2.06                                                                                    \\ \hline
\end{tabular}                                                           
\caption{\label{table:cut_flow_efficiency} Variation of signal and background efficiencies with cuts described in section~\ref{subsec:reconst}. The values in parentheses for backgrounds correspond to the different selection cuts for $M_h = 110$ GeV. The last column lists an estimate of the number of events with total integrated luminosity of 1000 fb$^{-1}$.  The number of signal events corresponding to $M_h = 70,~95~(110)$ GeV are given for benchmark cross sections ($\sigma_0$) of 1 (0.1) fb.}
\end{table}

\begin{figure}[h!]
    \centering
    \includegraphics[width=0.45\textwidth]{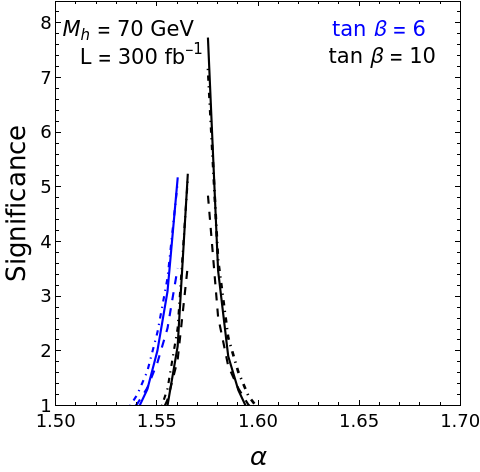}
    \includegraphics[width=0.45\textwidth]{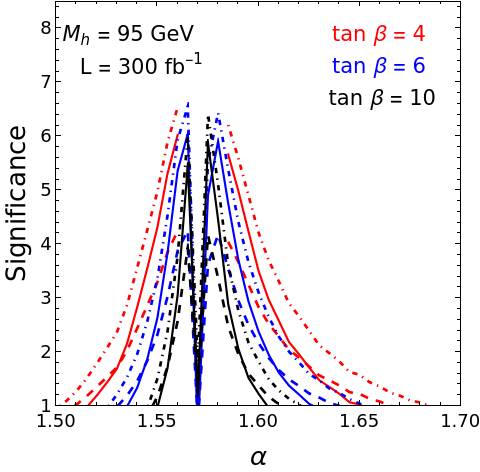}
  \includegraphics[width=0.45\textwidth]{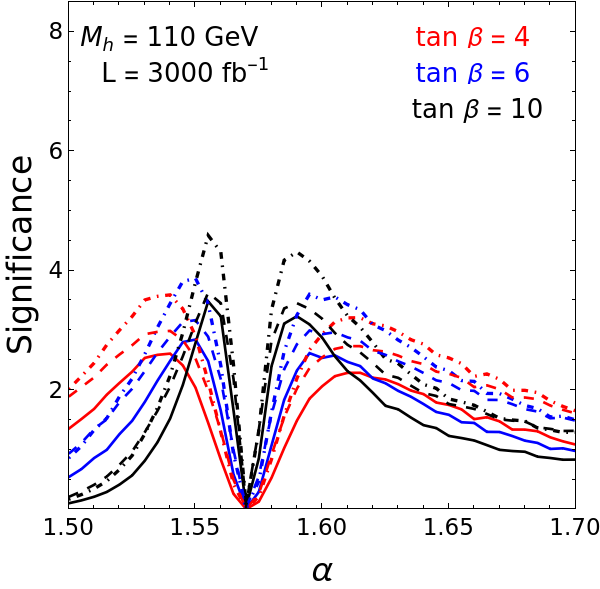}
    \caption{Significance of observing a light Higgs of mass = 70 GeV, 95 GeV and 110 GeV respectively for $M_{H^\pm} = 170$ GeV, and three values of $\tan \beta$. The dashed, dot-dashed and solid lines correspond to significances $S_1$, $S_2$ and $S_3$ respectively, as described in section~\ref{section:results}.
    \label{fig:significance}}
\end{figure}

In this section, we summarize the detection prospects of the light Higgs as well as charged Higgs in the channel $pp \to H^\pm h \to W^{\pm(*)}  h h \to \ell \nu_\ell \gamma \gamma b \overline{b}$ at 13.6 TeV LHC. The main results are shown in figure~\ref{fig:significance} 
in the plane of $\left(\alpha, \text{significance}\right)$. 
The scalar masses and $m_{12}$ parameters are fixed in accordance with the 
chosen benchmarks listed in Table~\ref{tab:benchmarks}.
The parameter $\alpha$ is varied in the range 1.5--1.7 and $\tan \beta$ is fixed to representative values of 4, 6 and 10.
The final states are selected according to the method described in section~\ref{section:collider_analysis}.
Our final state contains multiple BSM particles, therefore the significance of detection can be defined at each step of reconstruction.  Using $S/\sqrt{(S+B)}$ as the significance measure, we define significance $S_1$ after tagging of the light Higgs in the $\gamma \gamma$ decay mode. 
The significance $S_2$ is defined when both the light Higgses in the $\gamma \gamma$ as well as $b \overline{b}$ mode are identified. The significance $S_3$ is defined for the full final state {\it i.e},\,including the reconstruction of all new physics particles --- the charged Higgs in the $Wh$ mode and two light Higgses in the $\gamma\gamma$ and $b\overline{b}$ modes. Depending on the benchmark points, significance as a function of 
$\alpha$ exhibits striking features (see figure~\ref{fig:significance}) which we describe below:
\begin{itemize}
    \item The significances $S_1, S_2$ and $S_3$ decrease with the increase in the light Higgs mass for a fixed charged Higgs mass. This happens because of the phase space suppression in the charged Higgs decay to $Wh$. Due to this, the light Higgs of mass 70 and 95 GeV can be probed with 300 fb$^{-1}$ of the integrated luminosity whereas for $M_h = 110$ GeV, one would require higher values total integrated luminosity. 
    \item The significances are plotted only for the allowed values of the parameter space regions. The discontinuous regions in figure~\ref{fig:significance} for fixed value of $\tan\beta$ near the 
    fermiophobic limit correspond to parts of parameter space which are disallowed mainly the low mass diphoton constraint from ATLAS and CMS~\cite{ATLAS-CONF-2018-025, CMS:2018cyk}.  In addition, the stringent constraints from light Higgs search at LEP \cite{LEPHiggsWorkingGroup:2001opo} rules out low $\tan \beta$ for lighter Higgs masses. This explains the absence of $\tan \beta = 4$ line in the significance plot for $M_h = 70$ GeV in figure~\ref{fig:significance} where the limit requires $\tan \beta \gtrsim 5$.
    
    \item For $M_h =$ 95 and 110 GeV, the significance of tagging di-Higgs pair i.e. $S_2$ is larger than the significance of observing light Higgs in the diphoton mode i.e. $S_1$. For $M_h = 70$ GeV, the behavior is opposite i.e. $S_1$ is larger than $S_2$.  This happens because the distribution of b-jet $p_T$ peaks at higher values for $M_h = 95, 110$ GeV in comparison to $M_h = 70$ GeV which results in improved $b$-tag efficiency and consequently better significance for the final states.  The search for 70 GeV light Higgs might benefit significantly from further optimisation with techniques like BDTs.  
    
    \item The significance for all the benchmark points vanishes at the fermiophobic point. This is due to the vanishing coupling of the $h \to b \bar{b}$ decay mode at this point. Thus our significance displays a complementary behavior to the searches in the $4 \gamma$
    modes~\cite{Arhrib:2017wmo, Wang:2021pxc} where the significance peaks at $\alpha \to \pi/2$.

    \item For a fixed value of $\tan \beta$ in figure ~\ref{fig:significance}, the significances ($S_1$, $S_2$ and $S_3$) towards the left side of the fermiophobic point are slightly larger than the right hand side. This happens because of the dependence of production cross-section of $H^\pm h$ on $\cos(\beta-\alpha)$. Through Taylor expansion of the cross-section around $\alpha \to \pi/2$, we can see that the cross-section is higher for $\alpha < \pi/2$ in comparison to $\alpha > \pi/2$.  Consequently the significance of detection also follows the similar behavior. 
    
    \item For a given benchmark, the location of maximal significance changes with the choice of $\tan \beta$. 
    This results from the interplay between the branching ratio of Higgs to diphoton and $b\bar{b}$. The maximum significance occurs at the regions where the product of branching ratio of Higgs to $\gamma \gamma$ and $b \bar{b}$ attains the largest value. Upon increasing $\tan \beta$, the peak shifts away from the fermiophobic region which can be explained due to dependence of $h \to b\bar{b}$ coupling on $\cos{\alpha}/\sin{\beta}$.     
\end{itemize}

\section{Summary and conclusions} \label{section:conclusions}
We study the detection prospects of the charged Higgs as 
well as a fermiophobic Higgs pair at 13.6 TeV run of the LHC.  Most existing searches for charged Higgses rely on fermionic production and decay modes.  However, in Type-1 2HDM, the fermionic modes are naturally suppressed for large $\tan \beta$ and low $\tan \beta$ is already well-constrained from LEP and flavour constraints.  Therefore, there remains a gap in the search strategy that needs to be filled and searches in the bosonic modes at colliders have become necessary.   

We perform the search of charged Higgs where both production 
and decay is mediated by $H^\pm W^\mp h$ coupling. This helps us
to probe the regions with $\tan \beta \gg 1$. We analyze charged Higgs production via
$p p \to H^\pm h \to W^{\pm (*)} h h \to \ell \nu_\ell \gamma \gamma b \bar{b}$ for LHC run-3 and future HL-LHC runs.
In our analysis, we consider di-Higgs decays to $\gamma \gamma b\bar{b}$ in contrast with
the usual $4\gamma$ search popular in the literature which only works only very close to the exact fermiophobic limit.  
The two final states provide complementary information and must be probed together to provide full coverage of parameter space.  A similar final state can arise due to $H^\pm W^\mp A$ coupling. However, this process is subdominant when masses of the charged and the pseudoscalar Higgses are the same (as required by $\rho$-parameter constraints).

In order to reconstruct the charged Higgs in this bosonic decay mode, both neutral Higgses need to be identified.  Our analysis first tags light Higgs in the $\gamma \gamma$ mode followed by the second in $b\bar{b}$ channel.  Finally to reconstruct the charged Higgs mass, one needs to distinguish between the light Higgs emerging from the decay of the charged Higgs  i.e. $H^\pm \to W^{\pm(*)} h$ versus light Higgs produced in association with the charged Higgs i.e. $p p \to H^\pm h$. To resolve this ambiguity, we reconstruct the cluster mass variable of the charged Higgs along with the light Higgs boson utilizing two combinations of the final state particles i.e. $ H^\pm \to W^{\pm(*)} h \to \ell \nu_\ell \gamma \gamma $ and 
$ H^\pm \to W^{\pm(*)} h \to \ell \nu_\ell b \bar{b} $.  Imposing a suitable cut on cluster transverse
mass variable leads to the final reconstruction of the charged Higgs. 

To gauge improvement from each reconstruction, we define signal significance at each stage. $S_1$ is the significance after tagging
one light Higgs in the $\gamma \gamma$ mode, $S_2$ is defined after tagging di-Higgs production in the $\gamma\gamma b\bar{b}$
mode. $S_3$ is defined after reconstruction of the charged Higgs.  Our cut-based analysis finds that the substantial region 
of parameter space near the fermiophobic limit for the choice $M_{H^\pm} = M_A$ 
can be probed at 13.6 TeV run of LHC with total integrated luminosity varying between 300-3000 fb$^{-1}$. 
To summarise, our analysis displays that probing light charged as well as light nearly fermiophobic Higgs
is extremely promising at the future runs of the LHC.

We conclude by mentioning that there are several directions in which our analysis can be extended. 
For example, there still exists an unconstrained region in the parameter space where both CP-even and CP-odd Higgses can 
be lighter than the observed 125 GeV resonance.  In this case, depending on the decays of $h$ and $A$, 
the final state governed by the couplings viz., $H^\pm W^\pm h$ and $H^\pm W^\mp A$ can be similar. 
Thus, the reconstruction here gets even more complicated as it requires careful tagging of $h$, $A$ and finally $H^\pm$. 
We leave this study for our future work. Obviously, the analysis can be improved by using multivariate 
techniques instead of a simple cut-and-count analysis.  However, we hope this study shows an interesting channel possibly missed by experiments looking for di-Higgs searches in $\gamma \gamma bb$ that insist only on 125 GeV or heavier Higgses. 

\section*{Acknowledgements}

We would like to acknowledge Indian Association for the Cultivation of Science for providing help with computational resources. In addition, we would like to thank  Vinaykrishnan MB for his help with Delphes. SD would like to acknowledge Prof. Stefano Moretti for helpful discussions regarding formulation of the project in the initial phase. DB would like to acknowledge the fellowship support from FAPESP under contract 2022/04399-4 and would like to thank Institute of Mathematical Sciences for providing computational resources in the initial phase of the project. ND would like to acknowledge the support from the Department of Science and Technology through the Ramanujan grant SB/S2/RJN-070.

\appendix
\section{Fixing $m_{12}$ for different $M_h$ values} \label{append:m_12_values}
The Type-I 2HDM has seven independent parameters which we choose to be following: $M_H$, $M_h$, $M_{H^\pm}$, $M_A$,
$\alpha$, $\beta$ and $m_{12}$.  With the identification of ${H}$ with the observed 125 GeV resonance
and working in the limit, where $M_{H^\pm} = M_A$, five independent 
parameters remain. Among them, the mass of light and charged Higgses as well as the mixing angles affect the detectibility in our chosen channel. 
However, the parameter $m_{12}$ affects allowed values of $\tan \beta$  and therefore, indirectly affects ranges of production cross section and decay rates.  Therefore, we want to it fix it to an appropriate value. 

\begin{figure}[h!]
    \centering
    \includegraphics[width=0.3\textwidth]{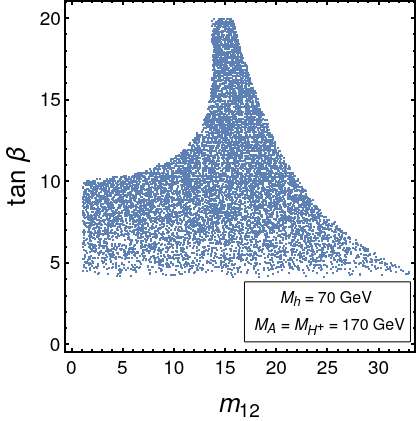}
    \includegraphics[width=0.3\textwidth]{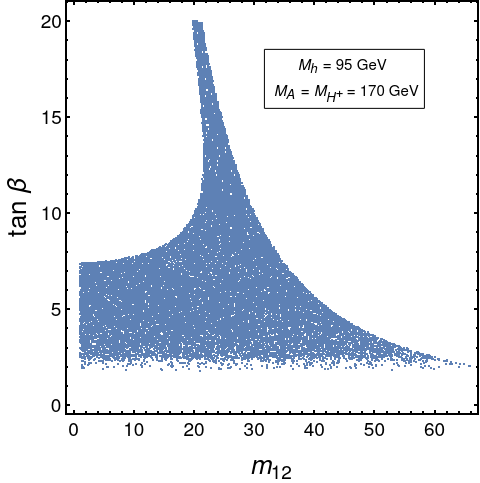}
    \includegraphics[width=0.3\textwidth]{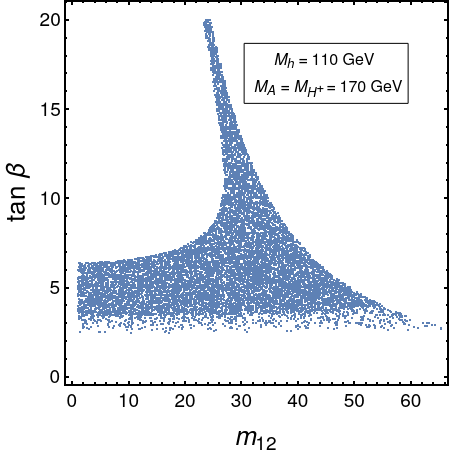}
    \caption{Allowed regions in the $(m_{12}, \tan \beta)$ plane for $M_h$ values 70 GeV, 95 GeV and 110 GeV respectively.}
    \label{fig:m12_vs_tan_beta_all_variation}
\end{figure}

For fixed values of $M_h$ and $M_{H^\pm}$, we see a sharp dependence of $m_{12}$ on $\tan \beta$, as can be seen in figure~\ref{fig:m12_vs_tan_beta_all_variation}.  This arises because of the positivity constraint 
on the scalar quartic coupling $\lambda_1$ given as:
\begin{equation}
    \lambda_1 =  \frac{M_H^2 \cos^2{\alpha} + M_h^2 \sin^2{\alpha}  -  m_{12}^2 \tan{\beta}}{v^2 \cos^2{\beta}}\;.
    \label{eqn:lambda1}
\end{equation}
On the other hand, there is no strong dependence of $m_{12}$ on $\alpha$.  Owing to mild dependence of our signal process on $m_{12}$, we determine 
a suitable value of it which allows for maximum allowed range in $\tan \beta$
for each given benchmark. Our selected values of $m_{12}$ are:
\begin{itemize}
\item For $M_h = 70$ GeV and $M_{H^\pm} = 170$ GeV, we fix $m_{12} = 15$ GeV.
\item For $M_h = 95$ GeV and $M_{H^\pm} = 170$ GeV, we fix $m_{12} = 25$ GeV.
\item For $M_h = 110$ GeV and $M_{H^\pm} = 170$ GeV, we fix $m_{12} = 32$ GeV. 
\end{itemize}

\bibliographystyle{jhep}
\bibliography{typ1}
\end{document}